\documentstyle[12pt,aaspp4]{article}
\lefthead{Kelsall et al. 1998}
\righthead{DIRBE Model of the IPD Cloud}
\begin{document}

%
%

\def\nWm2sr{~{\rm nW~m}^{-2}~{\rm sr}^{-1}}
\def\cf{{\it cf.}}
\def\eg{{\it e.g.}}
\def\ie{{\it i.e.}}
\def\etal{{\it et al.}}
\def\etals{{\it et al.\ }}
\def\wisk#1{\ifmmode{#1}\else{$#1$}\fi}
\def\lt     {\wisk{<}}
\def\gt     {\wisk{>}}
\def\le     {\wisk{_<\atop^=}}
\def\ge     {\wisk{_>\atop^=}}
\def\lsim   {\wisk{_<\atop^{\sim}}}
\def\gsim   {\wisk{_>\atop^{\sim}}}
\def\kms    {\wisk{{\rm ~km~s^{-1}}}}
\def\Lsun   {\wisk{{\rm L_\odot}}}
\def\Msun   {\wisk{{\rm M_\odot}}}
\def\ums     {\wisk{{\rm \mu m\ }}}
\def\um     {\wisk{{\rm \mu m}}}
\def\sig    {\wisk{\sigma}}
\def\bsl    {\wisk{\backslash}}
\def\by     {\wisk{\times}}
\def\amin   {\wisk{^\prime\ }}
\def\asec   {\wisk{^{\prime\prime}\ }}
\def\cc     {\wisk{{\rm cm^{-3}\ }}}
\def\deg    {\wisk{^\circ}}
\def\ddeg   {\wisk{{\rlap.}^\circ}}
\def\damin  {\wisk{{\rlap.}^\prime}}
\def\dasec  {\wisk{{\rlap.}^{\prime\prime}}}
\def\approxeq{$\sim \over =$}
\def\abouteq{$\sim \over -$}
\def\percm{cm$^{-1}$}
\def\percmsq{cm$^{-2}$}
\def\percmcub{cm$^{-3}$}
\def\perhz{Hz$^{-1}$}
\def\perpc{$\rm pc^{-1}$}
\def\persec{s$^{-1}$}
\def\peryr{yr$^{-1}$}
\def\te{$\rm T_e$}
\def\tenup#1{10$^{#1}$}
\def\to{\wisk{\rightarrow}}
\def\thin{\thinspace}
\def\uk{$\rm \mu K$}
\def\p{\vskip 13pt}
\def\etal{{\rm~et~al.~}}
\title{The COBE Diffuse Infrared Background Experiment Search
       for the Cosmic Infrared Background: II. Model of the
       Interplanetary Dust Cloud}
\author{T.~Kelsall, \altaffilmark{1,2}
        J.~L.~Weiland,\altaffilmark{3}
        B.~A.~Franz,\altaffilmark{4}
        W.~T.~Reach,\altaffilmark{5}
        R.~G.~Arendt,\altaffilmark{3}
        E.~Dwek,\altaffilmark{1}
        H.~T.~Freudenreich,\altaffilmark{3}
        M.~G.~Hauser,\altaffilmark{6}
        S.~H.~Moseley,\altaffilmark{1}
        N.~P.~Odegard,\altaffilmark{3}
        R.~F.~Silverberg,\altaffilmark{1}
        and E.~L.~Wright\altaffilmark{7}
        }

\altaffiltext{1}{Code 685, NASA GSFC, Greenbelt, MD 20771}
\altaffiltext{2}{email: kelsall@stars.gsfc.nasa.gov}
\altaffiltext{3}{Raytheon STX, Code 685, NASA GSFC, Greenbelt, MD 20771} 
\altaffiltext{4}{General Sciences Corp., Code 970.2, NASA GSFC, Greenbelt,
                 MD 20771}
\altaffiltext{5}{California Institute of Technology, IPAC/JPL, MS 100-22, 
          Pasadena, CA 91125}
\altaffiltext{6}{Space Telescope Science Institute, 3700 San Martin Drive,
                 Baltimore, MD 21218}
\altaffiltext{7}{UCLA Dept. of Astronomy, Los Angeles, CA 90024}

\slugcomment{Last revision: \today}

\begin{abstract}
\rightskip=0pt

The {\it COBE} Diffuse Infrared Background Experiment (DIRBE) was
designed to search for the cosmic infrared background (CIB) radiation.
For an observer confined to the inner solar system, scattered light
and thermal emission from the interplanetary dust (IPD) are major
contributors to the diffuse sky brightness at most infrared
wavelengths.  Accurate removal of this zodiacal light foreground is a
necessary step toward a direct measurement of the CIB.

The zodiacal light foreground contribution in each of 
the 10 DIRBE wavelength bands ranging from 1.25 to 240
$\mu$m is distinguished by its apparent seasonal variation over the
whole sky.  This contribution has been extracted by fitting the
brightness calculated from a parameterized physical model to the time
variation of the all-sky DIRBE measurements over 10 months of
liquid-He-cooled observations.  The model brightness is evaluated as the
integral along the line of sight of the product of a source function
and a three-dimensional dust density distribution function.  The dust
density distribution is composed of multiple components: a smooth
cloud, three asteroidal dust bands, and a circumsolar ring near 1 A.U.
By using a directly measurable quantity which relates only to the IPD
cloud, we exclude other contributors to the sky brightness from the
IPD model.

Using the IPD model described here, high-quality maps of the infrared
sky with the zodiacal foreground removed have been generated.
Imperfections in the model reveal themselves as low-level systematic
artifacts in the residual maps which correlate with components of the
IPD.  The most evident of these artifacts are located near the
ecliptic plane in the mid-infrared, and are less than 2\% of the
zodiacal foreground brightness.  Uncertainties associated with the
model are discussed, including implications for the CIB search.

\end{abstract}

\section{INTRODUCTION}
\rightskip=0pt

The effort to understand the nature of the ``zodiacal cloud'' or 
interplanetary dust (IPD) cloud is of long standing.  The first 
hypothesis for the cause of the zodiacal light was formulated over 
three centuries ago by Cassini (1685).  He proposed on the basis of 
careful visual observations that the brightness pattern seen in the 
night sky could be caused by a lenticular cloud of dust centered on 
the Sun with its main axis lying in the ecliptic plane.  This was an 
amazingly astute conclusion.  Though much later in formulation, 
another insightful interpretation of the structure of the IPD cloud was 
forwarded by Fessenkov in the early 1940's (Struve 1943).  
Fessenkov considered the IPD distribution as a prolate spheroid 
surrounded by a dust torus formed from the fragmentation of the 
asteroids in the asteroid belt.  Though even these early efforts came to 
reasonable conclusions as to the general shape of the IPD cloud, more 
recent data, such as those from the {\it IRAS} satellite and those 
from the {\it COBE} Diffuse Infrared Background Experiment (DIRBE), 
show the cloud structure to be rather more complex.  The primary objective
of the DIRBE, a search for the extragalactic infrared background, requires 
formulation of a detailed description of the IPD cloud so that its 
contribution to the sky brightness can be modelled and accurately
removed from measurements of the diffuse sky brightness.  The aim of
this paper is to describe the model developed by the DIRBE team.

A wide variety of evidence indicates that within a few 
astronomical units (AU) of the Sun the solar system is filled with 
dust of cometary and asteroidal origin.  The IPD reveals itself as a
diffuse component of the sky brightness, attributed to the scattering of
sunlight in the UV, optical and near-infrared, and the thermal
re-radiation of absorbed energy in the mid- and far-infrared.  At
infrared wavelengths from approximately $1-100~\mu$m, the signal from
the IPD is a major contributor to the diffuse sky brightness, and
dominates the mid-infrared ($\sim10-60~\mu$m) sky in nearly all
directions except very low Galactic latitudes.  With the aid of
mid-infrared photometric measurements, our picture of the zodiacal
cloud has evolved over the last fifteen years from a relatively simple
smooth distribution of dust to one of increasing complexity.  The
terms ``zodiacal cloud'' and ``IPD cloud'' now encompass several distinct
components, each of which may possess different grain properties and
experience different orbital dynamics (Dermott \etal 1996; Leinert \etal 
1998).  In addition to a smooth background distribution arising from a mix of
dust associated with asteroidal and cometary debris, there are also
contributions from smaller scale structures.  The {\it IRAS} data showed, 
and the DIRBE data confirm, that imposed on the brightness of the main 
cloud there are contributions from asteroidal dust bands 
(Low \etal 1984; Spiesman \etal 1995 and references therein).  More 
recently, a circumsolar ring theoretically proposed to arise from dust 
spiralling in from the outer solar system and being resonantly trapped 
by the Earth in orbits near 1 AU (Gold 1975; Jackson and Zook 1989; 
Marzari and Vanzani 1994a, 1994b; Dermott \etal 1994) was compellingly 
confirmed by Reach et al. (1995) through use of a simple IPD model 
and the DIRBE data. Finer scale features such as dust trails in cometary 
orbits have also been observed in {\it IRAS} data (Sykes \etal 1986).

The issue as to the ultimate source of the dust is still open, but 
satellite observations of the sky brightness and in situ 
measurements of dust particles are advancing our understanding of the 
relative contributions from comets, asteroids, and the
interstellar medium.  From a study of 200 captured dust particles, 
Schramm, Brownlee, and Wheelock (1989) find that it is possible to
ascribe 45\% of the particles to cometary origin and 37\% to asteroidal
origin.  Using dynamical analysis to construct the configuration of 
the IPD cloud from comets and asteroids, Liou, Dermott,
and Xu (1995) found a match to the {\it IRAS} brightness profiles 
using a mixture of 74\% cometary and 26\% asteroidal dust. 
More recently, using numerical modelling to determine the evolution of
dust from asteroidal collisions, Durda and Dermott (1997) find that
34\% of the IPD could arise from collisional destruction of asteroidal
family members (10\%) and main-belt members (24\%).  In addition to the
dust formed from members of the solar system, Baguhl \etal (1995) find
in the Ulysses satellite dust-detection data a distinct signature in
the impact directions indicating that interstellar dust is flowing
into the solar system.  The contribution to the brightness of the IPD
from the interstellar dust is small, as shown by Grogan, Dermott, and
Gustafson (1996), and it is not considered in the modelling described
in this paper. Though there have been advances in estimating the source of the
dust, reality appears to be yet more complex.  Divine (1993) has shown
from an interpretation of ground-based and satellite measurements that
the IPD is best represented as being composed of five distinct
dust-family components.  On the basis of this work, Staubach, Divine \& 
Gr\"{u}n (1993) have estimated the sky brightness arising from 
the five-component cloud.  Clearly in the future there will be a
strong symbiotic relationship between the modelling of the brightness
of the IPD and the growing intricate and precise theoretical and
observational studies of its nature.

Data from {\it IRAS} (Neugebauer \etal 1984; Beichman 1987) have been
instrumental in advancing studies of the infrared zodiacal emission.
However, the {\it IRAS} database does have some limitations, including
relatively uncertain calibration of the photometric zero point, wavelength 
coverage only from 12 to 100 $\mu$m,  and sparse solar elongation 
sampling of most celestial directions.  Diffuse infrared
sky brightness measurements have been greatly extended by the Diffuse
Infrared Background Experiment (DIRBE), launched as part of
the Cosmic Background Explorer ({\it COBE}) in 1989.  The DIRBE is an
absolutely calibrated photometer with an instrumentally-established
photometric zero point, designed to survey the sky in ten broad
photometric bands from $1.25-240~\mu$m.  As described in more detail
in \S~2, the unique DIRBE scanning strategy and dense, continuous
sampling of the sky at solar elongations from $64^\circ-124^\circ$
over a 10-month interval have provided a previously unequaled infrared
photometric database from which to study the IPD in both scattering
and thermal wavelength regimes.

A major thrust in photometric studies of the IPD is the construction
of models which can be used to reproduce the observed brightness
distribution.  The development of such models is not always motivated
by a wish to understand the IPD for itself, but rather as a means for
removing an obfuscating foreground which masks contributions from
Galactic and extragalactic sources.  The focus of this paper is to
describe the model for the IPD which has been used to remove the
zodiacal foreground from each of the ten DIRBE bands to permit
investigation of the residual sky brightness for evidence of the
cosmic infrared background (CIB; defined here as all diffuse infrared
radiation of extragalactic origin).  Since it was anticipated that the
CIB may be faint compared to the IPD and Galactic foregrounds, the aim
has been to generate an IPD model which could reproduce the zodiacal
brightness to within a few percent or better.  

A number of models and
modelling techniques have been developed to describe the infrared
zodiacal light (e.g., Boulanger \& Perault 1988; Jones \& Rowan-Robinson 1993; 
Wheelock \etal 1994; Dermott \etal 1996).  All models, including the
one described here, have limitations brought about primarily because
of the general lack of detailed information on IPD grain properties,
populations, and spatial distribution.  No previous modelling effort, 
however, has been designed to take advantage of the spectral, temporal
and angular coverage which DIRBE provides.  

The remainder of this paper describes the development of a 
parameterized physical model of the
IPD in which the parameters are determined by requiring the model
brightness to match the measured time-dependence of the sky brightness 
in fixed celestial directions over the whole sky.  Section~2 briefly 
describes the DIRBE instrument, scan strategy, and data 
reduction methods.  In \S~3, the
motivation behind the chosen modelling approach is discussed.  The 
details of the model itself are presented in \S~4. Results of the
modelling, including uncertainties, are presented in \S~5.  Section~6
contains a discussion of the implications of the model uncertainties
for the CIB search, and conclusions are presented in \S~7. 

This paper is part of a series outlining the analysis of DIRBE data in
the search for the cosmic infrared background.  Paper I (Hauser \etal
1998) reports the observational results of the DIRBE search, compares 
them with previous work, and briefly discusses their implications.   The
foreground removal techniques used in the DIRBE search are described 
in two separate papers: this paper (Paper II), and 
Paper III (Arendt \etal 1998), which describes the Galactic foreground 
removal procedures and summarizes the systematic uncertainties arising from 
the total foreground discrimination process.  A further discussion of the 
cosmological implications of the results described in Paper I is presented by 
Dwek \etal (1998,  Paper IV).

\section{DIRBE INSTRUMENT AND DATA OVERVIEW}

The data used in this analysis are broad-band photometric measurements 
obtained with the DIRBE instrument during its 10 months of cryogenic
operation (1989 Dec 11 to 1990 Sep 21) aboard the Cosmic Background
Explorer ({\it COBE}) satellite.  A detailed description of the {\it
COBE} mission has been given by Boggess \etal (1992), and the DIRBE
instrument has been described by Silverberg \etal (1993).  The DIRBE
provided a simultaneous survey of the sky in 10 photometric bands at
1.25, 2.2, 3.5, 4.9, 12, 25, 60, 100, 140 and 240 $\mu$m using a
$0.7^\circ\times0.7^\circ$ field of view.  Linear polarization data were 
gathered in the three shortest wavelength bands, but those data are 
not used in the model described here.

A stable DIRBE photometric system was established using a combination of 
internal and celestial sources ({\it COBE} DIRBE Explanatory
Supplement 1997).  The short-term instrumental gain stability, during the 
ten months of cryogenic operation at 
1.8 K, was maintained using observations of a stable internal 
stimulator $\sim$6 times per orbit.  The long-term gain stability 
was monitored using 
bright stable celestial sources observed during normal sky scans, 
which were placed on an instrument-unique relative photometric 
system.  The combination of using both the internal stimulators and 
celestial sources achieved a relative photometric system that was stable 
to $\sim1\%$ over the cryogenic mission at most wavelengths.  

The data at 100 $\mu$m, however, had a special photometric problem.  
The Ge:Ga detector used at these wavelengths exhibited a gain 
instability referred to as 
``photon induced responsivity enhancement'' (PIRE; {\it COBE} DIRBE 
Explanatory Supplement 1997).   The PIRE effect is a form of hysteresis 
evidenced by an increase in responsivity when the DIRBE beam swept 
across a bright extended region, such as the inner Galaxy and
the Cygnus spiral arm. The brightness recorded in pixels in the vicinity of 
these regions depended on the direction from which they were 
approached (i.e., on the brightness of the pixels previously scanned), with 
differences at 100 $\mu$m as large as 25\%.  A 
non-linear correction was developed to mitigate the PIRE
effect on the data.  This correction significantly reduced the effect, leaving 
residual scan-dependent differences of the order of 5\% or less,
though in the brightest Galactic regions differences can still be as
large as 10\%.  In contrast, the Ge:Ga detector used at 60~$\mu$m, for
which a PIRE correction was not implemented, was better-behaved,
registering a difference of no more than 3\%. Since the direction from
which each pixel was approached in the survey varied in a systematic 
way with time, the PIRE effect may have created a spurious apparent 
temporal modulation of the brightness.  Such modulation would not be 
coherent over a span of more than a few degrees, and would occur only 
at low Galactic latitudes.  Since our search for the CIB excluded regions
at low Galactic latitude (Paper I), this does not pose a serious problem 
for the DIRBE investigation.

Another essential feature of the calibration was the monitoring of 
the shape and centroids of the beams for each detector.  This was done 
by assembling the transit data from selected bright celestial sources.
The location of any transit was determined to the order of 1.5 arcminutes, 
which is a very small fraction of the 42-arcmin DIRBE beam.  We found 
no solid angle variations in any band during the 10 months of cryogenic
operation.  Any variation must be smaller than the accuracy with 
which we were able to measure the solid angle, which was better 
than 1\% from 1.25 to 25 $\mu$m, 
and better than 5\% from 60 to 240~$\mu$m. These
numbers include the uncertainty due to the possible contribution of
stray light.  The month-to-month beam centroid peak-to-peak variations
during the cold mission were at most 1 arcminute.  Errors in the
centroid correction for wavelengths from 1.25 to 4.9 $\mu$m are certainly
much less than this because the reference beams for these bands were
formed during intervals in which the variation about the mean was much
less than 1 arcminute.  Beam centroid errors for the longer wavelength 
bands, for which a single reference beam was used for the entire cold 
mission, are estimated to be less than 0.5 arcminute. 

The absolute gain calibration at each wavelength was based upon
observations of a single well-calibrated celestial object.  The
selected calibration objects were Sirius (1.25--12~$\mu$m), NGC 7027
(25~$\mu$m), Uranus (60 and 100~$\mu$m), and Jupiter (140 and
240~$\mu$m).  The uncertainty in the absolute gain calibration is
$\sim3-5\%$ at $1.25-12~\mu$m and $\sim10-15\%$ at $25-240~\mu$m
(Paper~I).  The DIRBE intensity measurements are reported in units of
MJy~sr$^{-1}$, quoted at the ten nominal wavelengths with effective
bandwidths calculated assuming the observed source energy distribution
is $\nu I_{\nu}={\rm constant}$.  This necessitates the use of a color
correction when computing the zodiacal light contribution (\S~4).

A critically important part of the DIRBE calibration, and one which
distinguishes the DIRBE from most other infrared instruments, was
establishing a true zero point for the photometric system.  This was
accomplished instrumentally.  The DIRBE optical system was a totally
off-axis design with multiple field and pupil stops which limited
external stray light to levels well-below the sensitivity
limit for the measurements at all wavelengths, $\nu I_{\nu}<1 \nWm2sr$.  
The DIRBE instrument was equipped with a cold chopper, which modulated the
beam at 32~Hz, and a cold shutter located at a field stop which could
completely shut out sky light from reaching the detectors.  When the
shutter was open, each detector measured the difference between the
sky brightness and that of an internal beam stop maintained at a
temperature below 3~K so as to emit essentially no flux at the DIRBE
wavelengths.  The shutter was closed about six times per orbit and the
instrumental offset signal at 32 Hz was measured.  This offset was
stable throughout the entire 10 month cryogenic mission, and was
subtracted from the data obtained while scanning the sky.  A real
radiative instrumental offset was present, largely due to emission
from the JFETs used in the signal amplifiers for the detectors which
were mounted in the detector assemblies.  This offset, which was
significant only at wavelengths of 60 $\mu$m and longer, was confirmed
by turning off the JFETs sequentially and measuring the change in
offset in the other detectors.  The offset uncertainty was carefully
assessed using special on-orbit tests, which showed that the size of
the offset did not depend on whether the shutter was open or closed.
The zero-point uncertainty, which is substantially below the sky
brightness at all wavelengths, does not affect the IPD model, but does
affect the systematic uncertainty in the CIB at long wavelengths
(Paper I).

The observing strategy for DIRBE was designed to monitor the change in
the sky brightness contribution from the IPD toward each celestial
line of sight as a function of time, thus providing a richer infrared
database for studying the IPD cloud than previously available.  Like
{\it IRAS}, the {\it COBE} spacecraft was placed in a Sun-synchronous
polar orbit at 900 km altitude, but, unlike
{\it IRAS}, the spacecraft also rotated (rate $\sim0.8$~rpm).  The
{\it COBE} spin axis was nominally held fixed at an angle of 94\deg\
from the Sun, with the DIRBE instrument line-of-sight angled 30\deg\
from the spin axis.  The 30\deg\ cant caused the DIRBE to execute a
cycloidal track on the sky as the {\it COBE} both rotated and orbited
the Earth.  With a half-angle of 30\deg, the viewing swath enclosing
this track covered one half of the sky after one orbit of the
spacecraft (Fig. 1a).  Because the field-of-view of the DIRBE was
$0.7^\circ\times0.7^\circ$, and the {\it COBE} orbital velocity was
about 3.5\deg/minute, the one-orbit scan swath had gaps in coverage. 
Full sampling of one-half of the sky was achieved after about 1 day of
observations (Fig~1b); the highest density of observations occured
along the edges of the viewing swath, and the lowest density at the
center of the swath.  It also follows from the scanning geometry that
the solar elongation angle (angle between the Sun and the
line-of-sight) at which an observation was made is dependent on the
location within the viewing swath: those positions along the swath
edges closest to the Sun were viewed at elongations $e~\sim
64$\deg, and those along the edges furthest from the Sun at
$e~\sim 124$\deg, with a smooth continuum between. 

The {\it COBE} orbit precessed one degree per day, so the DIRBE
viewing pattern on the sky shifted by this amount along the ecliptic
plane each day.  Complete sky coverage was thus achieved within four
months, with more uniform sampling in six months.  Figures 1c and 1d
show the average of all observations of the sky at 12~$\mu$m for
periods of one week and the full mission, respectively.  Because of
the 1\deg/day orbital precession, a fixed celestial location near the
ecliptic plane was observed at solar elongations from 64\deg\ to
124\deg\ over the course of 60 days each half year.  Above ecliptic
latitudes of $\sim60$\deg\, the amplitude of the full range in
elongation decreases about two degrees per degree of latitude, and is
zero directly at the ecliptic poles.  In this way, all accessible
solar elongation angles in the range 64\deg\ to 124\deg\ were sampled
for each location on the sky.

The photometric data from the DIRBE time-ordered scans described above
were converted into maps of the sky by associating each observation
with a pixel in the quadrilateralized spherical cube projection
(O'Neill \& Laubscher 1976; {\it COBE} DIRBE Explanatory Supplement
1997) in geocentric ecliptic coordinates, epoch 2000.  The map
resolution was chosen such that pixels are roughly 0.32\deg\ on a side, 
yielding 393,216 approximately equal-area pixels.  A set of forty-one 
Weekly Sky Maps was created for the period of cryogenic operation, 
in which all individual observations of pixels sampled during one week were
robustly averaged to produce a single average intensity per pixel at
each wavelength.  Pixels within $\sim 60$\deg\ of the ecliptic plane
typically have $\sim10-15$ samples per week over which to average; the
most densely sampled pixels near ecliptic latitudes $|\beta|=60$\deg\
have $\sim30$ samples per week. The averaging interval of one week
was chosen so as not to overly smear apparent temporal variations of
the zodiacal light (compare Figs. 1b and 1c), and yet provide a
reasonably compact database.  This set of Weekly Sky Maps is the basis
for this study; specifically the Pass 3b version of these products
released through the NSSDC in 1997. 

All analysis has been performed using data from the original sky-cube
coordinate system.  For illustrational purposes, maps shown in this
paper are reprojected into an ecliptic coordinate Mollweide
projection.  The Mollweide projection is an equal-area projection with
the convenient properties that longitudes are equally spaced at each
latitude and all lines of constant latitude are straight horizontal
lines (though unequally spaced).

\section{MOTIVATION AND MODELLING TECHNIQUE}

The development of the IPD cloud model was driven by requirements
imposed by the ultimate goal of determining the contribution of the
isotropic CIB signal in the infrared. Thus, the method described here
was designed to provide a means for accurately subtracting the
zodiacal signal from the DIRBE observations at 1.25 through 240~$\mu$m, 
while preserving any isotropic component of the sky brightness
unrelated to the IPD; i.e., our method includes no arbitrary 
zero-point constants. The scheme follows what is now a canonical 
approach adopted in early models (Haug 1958; Leinert \etal\ 1976) of 
representing the IPD cloud geometric and radiative 
characteristics in a parametric form.  This procedure has been
utilized in a number of recent investigations  (e.g., Murdock \& Price 1985;
Deul \& Wolstencroft 1988; Rowan-Robinson \etal 1990; Jones and
Rowan-Robinson 1993).  The physics of the IPD cloud is embedded in the
representations of the scattering and emissivity functions and the
form factors for the various components (i.e., main cloud, asteroidal
bands, and circumsolar ring).  

Figures~2a and 8a illustrate the nature
of the difficulty in generating such a model.  The observed infrared
sky brightness is a complex mix of foregrounds due to interplanetary
dust, starlight, and interstellar dust, as well as an extragalactic
background which is presumed to be isotropic.  The relative mixture of
foregrounds is wavelength dependent; only in the mid-infrared does
emission from the IPD contribute $90\%$ or more of the total sky
brightness.  Although the shape of the underlying zodiacal ``lower
envelope'' is clearly visible in the figures from $1.25-100~\mu$m,
spatially disentangling the Galaxy from the IPD signal with high
precision is challenging (Hauser 1988).  The problem is additionally
complicated by the presence of low-constrast, smaller scale structures
within the IPD cloud such as the dust bands and circumsolar 
ring (\S~4.2.2 and \S~4.2.3). 

In order to determine the contribution from interplanetary dust
without modifying any of the other contributions, the one
unique signature of the zodiacal light is used: it is the only component of
the sky brightness that is not fixed on the celestial sphere.  For an
Earth-based observer, the IPD brightness observed in a given celestial
direction on a given day depends on the observer's viewing aspect.
Changes in viewing aspect resulting from the DIRBE cycloidal scan
strategy and motion through the IPD cloud cause different path lengths
through the dust to be sampled (and thus different portions of the
dust density and temperature distributions), which gives rise to
apparent temporal variations in the brightness observed toward a fixed
celestial direction.  Figure~3 is a schematic representation of how
the zodiacal cloud signal is modulated in two directions.  Toward the
ecliptic pole (Fig.~3a), the main causes for variation of the
brightness are the motion of the Earth with respect to the inclined
midplane of the dust distribution, as well as the motion of the Earth
radially due to its orbital eccentricity.  At lower latitudes
(Fig.~3b), the apparent temporal variation is primarily due to the
changing solar elongation of the line of sight as the Earth orbits
about the Sun.

The technique used here exploits these effects to separate the
light scattered and emitted by the IPD from all other brightness 
contributions by imposing an
analytical form for the interplanetary dust density distribution,
thermal emission characteristics, and scattering phase function.  This
parameterized model is then optimized to match the observed temporal
variations in selected directions.  The formulation of the
parameterized model is described in \S~4.  Once the best-fit model
parameters are determined, the brightness of the zodiacal light is
then evaluated for all DIRBE weekly observations via a line-of-sight
integral of the three-dimensional model.

\section{PARAMETRIC IPD MODEL}

The DIRBE IPD model is a parameterized physical model whose
formulation is similar, but not identical, to that used in creating
the {\it IRAS} Sky Survey Atlas (Wheelock \etal 1994).  The DIRBE
model consists of the integral along the line of sight of the product
of a source function and a three-dimensional density distribution
function.  The emissivity function includes both thermal emission and
scattering terms.  The 3-dimensional dust density distribution is
composed of multiple components -- a smooth cloud, three dust bands,
and a circumsolar dust ring just beyond 1 AU.  Earlier versions of the
model have been described by Reach \etal (1996) and Franz \etal
(1996).

\subsection{The Brightness Integral}

The model for the IPD foreground computes the brightness of the
zodiacal light observed at wavelength $\lambda$ for each pixel~$p$ at
time~$t$ as the integral along the line-of-sight of scattered and
thermal emission contributions, summed over each density component
$c$:
\begin{equation}
Z_{\lambda}(p,t) = \sum_{c} \int n_{c}(X,Y,Z)[A_{c,\lambda}
F_\lambda^\odot \Phi_{\lambda}(\Theta) + (1-A_{c,\lambda})
E_{c,\lambda} B_{\lambda}(T) K_{\lambda}(T)]~ds,
\end{equation}
where $n_{c}(X,Y,Z)$ is the three-dimensional density for each of the
components, $A_{c,\lambda}$ is the albedo for component $c$ at
wavelength $\lambda$, $F_\lambda^\odot$ is the solar flux,
$\Phi_{\lambda}(\Theta)$ is the phase function at scattering angle
$\Theta$, $E_{c,\lambda}$ is an emissivity modification factor, which
measures deviations from the blackbody thermal radiance function
$B_{\lambda}(T)$, and $K_{\lambda}(T)$ is the DIRBE color-correction
factor appropriate for $B_{\lambda}(T)$.  The dust grain temperature
$T$ is assumed to vary with distance from the Sun as 
$T(R)=T_0 R^{-\delta}$. The derived model value of $\delta = 0.467$
is very close to the theoretical value of 0.5 for large grey grains in
radiative equilibrium.

As in the case of {\it IRAS} photometry, the DIRBE broad-band
photometric measurements $I_\lambda$ are quoted at fixed nominal
wavelengths with bandwidths determined assuming the source energy
distribution is $\lambda I_{\lambda} = {\rm constant}$.  Since the
model parameters are optimized by fitting to the DIRBE data, a color
correction $K_{\lambda}(T)$ must be applied to each evaluation of the
blackbody source term in order to compute the zodiacal brightness in a
fashion consistent with the DIRBE database; this is equivalent to
convolving the source function with the DIRBE spectral response.  A
color-correction of 1.0 is used for the scattering source function.
The color corrections are taken from those tabulated in the {\it COBE}
DIRBE Explanatory Supplement (1997).

The treatment of the albedos $A_{c,\lambda}$ and emissivity
modification factors $E_{c,\lambda}$ is important for this work.  The
aim is to achieve high accuracy in modelling the zodiacal light from
$1.25-240~\mu$m, but the accuracy of the absolute calibration from
wavelength to wavelength in the mid-infrared is relatively poor
(\S~2). Furthermore, it is unlikely that the assumption of a single
grain temperature at each distance $R$ is entirely accurate.  Models for
the emission from interplanetary dust predict that a small range of
temperatures contributes to the mid-infrared emission, expanding to a
wider range at shorter wavelengths (Reach 1988).  The
wavelength-dependence of the albedo is essentially unknown, and a
constant albedo is unlikely.  In order to allow for the imprecision of
the band-to-band calibration as well as the restrictive nature of the
spectral model, the factors $A_{c,\lambda}$ and $E_{c,\lambda}$ are
allowed to be free parameters.  Albedos are allowed to be non-zero
only at the shorter wavelengths (1.25, 2.2, \& 3.5~$\mu$m) and
$E_{c,\lambda}$ is set to 1.0 at 1.25 and 2.2~$\mu$m.  
The $E_{c,\lambda}$ is normalized to unity at 25~$\mu$m.  As coded,
each density component could have its own particle properties, as
delineated by $A_{c,\lambda}$ and $E_{c,\lambda}$.  In practice,
however, we allowed for three groups of $A_{c,\lambda}$ and
$E_{c,\lambda}$: one for the smooth cloud, one for all of the dust
bands, and one for the circumsolar ring.  In addition, at
$1.25-3.5~\mu$m, where the dust bands and circumsolar ring have a
small contribution to the cloud brightness, the ring and dust band
values for $A_{c,\lambda}$ and $E_{c,\lambda}$ were not optimized, but
rather assumed to be identical to the values found for the smooth
cloud at those wavelengths.

There is no definitive work which establishes a phase function in the
infrared.  Initial modelling attempts used the Henyey-Greenstein
formulation determined by Hong (1985) for visible data, which allowed
for no dependence of the phase function on wavelength.  The final
model incorporated a wavelength-dependent three-parameter functional
form which was capable of reproducing the shape of Hong's function,
but could also be optimized for the infrared.  The functional form of
the phase function $\Phi_{\lambda}(\Theta)$ is
\begin{equation}
\Phi_{\lambda}(\Theta) = N \left[ C_{0,\lambda} + C_{1,\lambda} \Theta + 
e^{C_{2,\lambda} \Theta} \right]
\end{equation} 
where $\Theta$ is the scattering angle in radians.  The three free 
parameters for each wavelength are $C_{0}$, $C_{1}$ and $C_{2}$.
The factor $N$ is not free, being a function of the three $C$ parameters 
and serving to normalize the phase function
such that the integral over $4\pi$ steradians is 1.

\subsection{Model Components and Geometry}

The model calculations are performed in heliocentric ecliptic
coordinates, $(X, Y, Z)$. The coordinate transformation for a grid
point at a distance $s$ from the Earth along a line-of-sight at
geocentric ecliptic coordinates $(\lambda, \beta)$ is
\begin{eqnarray}
X      & = & R_\oplus \cos \lambda_\oplus + s \cos\beta
\cos\lambda  \nonumber \\
Y      & = & R_\oplus \sin \lambda_\oplus + s \cos\beta
\sin\lambda \\
Z      & = & s \sin\beta \nonumber \\
R      & = & \sqrt{X^2 + Y^2 + Z^2} \nonumber 
\end{eqnarray}
where $R_\oplus$ is the Earth--Sun distance, and $\lambda_\oplus$ is
the heliocentric longitude of the Earth ($\lambda_\oplus=\pi -
\lambda_\odot$) on the date of observation.  The eccentricity of the
Earth's orbit was included in the model.

The integration along the line of sight was performed by
Gauss-Laguerre quadrature with 50 points.  For ecliptic latitudes
$|\beta|\leq20\deg$, Simpson's rule with $\sim 200$ steps was used in
order to more accurately evaluate density gradients in small-scale
structures closer to the ecliptic plane.  Integration along the line
of sight is performed from the Earth to an outer radial cutoff of 5.2
AU from the Sun, which roughly corresponds to the orbit of Jupiter.

All dust density components were assumed to be intrinsically
time-independent and to have a plane of symmetry (although not
necessarily the same plane in all cases).  The only exception to this 
is the trailing blob in the circumsolar ring, which follows the 
Earth in its orbit.  Isodensity contours of the total model density
and its components are shown in Figure~4. The parameterized form used
for each of these components is described in detail below. 

\subsubsection{Smooth Cloud}

The center of the smooth dust cloud was allowed to be offset from
the Sun by $(X_0, Y_0, Z_0)$, so that the cloud coordinates were
translated as follows:
\begin{eqnarray}
X^\prime  & = & X - X_0 \nonumber \\
Y^\prime  & = & X - Y_0 \\
Z^\prime  & = & Z - Z_0 \nonumber \\
R_{c}     & = & \sqrt{X^{\prime 2} + Y^{\prime 2} + Z^{\prime 2}}. \nonumber
\end{eqnarray}
The symmetry plane of the smooth cloud was also allowed to be tilted
with respect to the ecliptic plane, so that the vertical structure is
dictated by the height above the tilted midplane:
\begin{equation}
Z_c = X^\prime \sin\Omega \sin i - Y^\prime \cos\Omega\sin i +
Z^\prime\cos i, \label{eqtilt}
\end{equation}
where $i$ and $\Omega$ are the inclination and ascending node of the
midplane, respectively.

The density of the smooth cloud was presumed to be of a form which is
separable into radial and vertical terms:
\begin{equation}
{n_c(X,Y,Z)} = {n_0} R_{c}^{-\alpha} f(\zeta),
\end{equation}
where $\zeta\equiv |Z_{c}/R_{c}|$.  The separation into radial and
vertical terms is typical of several models for the cloud density in
the literature (e.g., as summarized by Giese \etal 1986).  However, it 
should be noted that there are some theoretical reservations as to its 
applicability to modelling the IPD cloud (Banaskiewicz, Fahr and 
Scherer 1994; Fahr, Scherer  and Banaskiewicz 1995). The radial 
power-law is motivated by the radial distribution
expected for particles under the influence of Poynting-Robertson drag,
which causes their orbital semi-major axes to decay such that their
equilibrium distribution is $1/R$. The assumption that the vertical
distribution depends only on $\zeta$ is motivated by the fact that
Poynting-Robertson drag does not affect the orbital inclination of
particles as they spiral into the Sun.  The radial power-law index,
$\alpha$, is a free parameter. 

The vertical distribution $f(\zeta)$ was written in a form
respresenting a widened, modified fan model:
\begin{equation}
f(\zeta) = e^{-\beta g^\gamma},
\end{equation}
\noindent where
\[ g = \left\{ \begin{array}{ll}
                \zeta^2/2\mu  & \mbox{for $\zeta < \mu$} \\ 
                \zeta-\mu/2   & \mbox{for $\zeta \geq \mu$}
               \end{array}
       \right. \]            
and $\beta$, $\gamma$ and $\mu$ are free parameters. (Note that
$\beta$ is not to be confused with the geocentric ecliptic latitude.)  
The vertical distribution is critically important in connecting the 
local density of the dust, which is very well determined by the DIRBE 
data, with the integrated column density.

This model is somewhat different from the one fitted to the {\it IRAS}
data in production of the {\it IRAS} Sky Survey Atlas  (Wheelock \etal\
1994), because we used a spherical radius rather than a cylindrical
one.  Furthermore, the function $g$, which replaces $\zeta$ in the
exponential of the {\it IRAS} model, rolls off at small values and
avoids the cusp in the midplane that is present in traditional fan
models.  Isodensity contours of the smooth cloud model are shown in
Figure 4b.

\subsubsection{Dust Bands}

The dust bands were discovered in the {\it IRAS} data (Low \etal
1984), and are believed to be asteroidal collisional debris (Dermott
\etal 1984; Sykes \etal 1989).  The dust bands have been studied using
the DIRBE data (Spiesman \etal 1995), confirming the observational
results from {\it IRAS} data and extending them to the near-infrared.
In particular, the parallactic and spectroscopic distances to the
bands are less than the distance to the asteroid belt, so that the
material producing them is likely to be debris spiralling into the Sun
under Poynting-Robertson drag.  Three band pairs are included, which
appear at ecliptic latitudes around $\pm1.4^\circ$, $\pm10^\circ$ and
$\pm15^\circ$ in the sky maps.  These have been attributed (Sykes
\etal 1989, Reach \etal 1997) to a blend of the Themis and Koronis
families ($\pm1.4^\circ$), the Eos asteroid family ($\pm10^\circ$),
and the Maria/Io family ($\pm15^\circ$).
All band pairs were centered on the Sun, but were allowed to be
inclined with respect to the ecliptic plane.  Each band pair $i$ had
its own inclination $i_{Bi}$ and ascending node $\Omega_{Bi}$.  A
transformation similar to that in equation~5 was used to define the
vertical distance from the midplane of band pair $i$, $z_{Bi}$.

For this work, a dust band density based on the migrating model (Reach
1992) was used, but with a simpler analytic formulation that is easier
to evaluate and optimize.  A modification was added in the form of a
multiplicative factor which allowed for only ``partial'' migration,
i.e., a cut-off at a minimum radius:
\begin{equation}
n_{Bi}(X,Y,Z) = {3n_{3Bi}\over R}
\exp\left[-\left({\zeta_{Bi}\over\delta_{\zeta_{Bi}}}\right)^6\right]
\left[v_{Bi}+\left({\zeta_{Bi}\over\delta_{\zeta_{Bi}}}\right)^{p_{Bi}}
\right]
\left\{1-\exp\left[-\left({R\over\delta_{{R}_{Bi}}}\right)^{20}\right]\right\},
\end{equation}
where $n_{3Bi}$ is the density at 3 AU of band $i$, $\zeta_{Bi}\equiv
|z_{Bi}/R_c|$, and $\delta_{\zeta_{Bi}},v_{Bi}$ and $p_{Bi}$ are
adjustable shape parameters.  The parameter $\delta_{{R}_{Bi}}$ determines
the distance to which band $i$ migrates in toward the Sun.

\subsubsection{Circumsolar Ring}

The Earth temporarily traps migrating dust particles into resonant
orbits near 1 AU if they are in low-eccentricity orbits, as is
expected for asteroidal debris (Jackson and Zook 1989; Marzari and 
Vanzani 1994a, 1994b; Dermott \etal\ 1994; Dermott 1996).
We have confirmed the existence of the dust ring near 1 AU by
subtracting a smooth cloud model from two weekly sky maps, revealing
the signature of the ring in remarkable agreement with the predictions
(Reach \etal 1995).  For the DIRBE IPD model, an empirical ring
density function was developed to emulate the numerical simulations of
Dermott \etal (1994).  It consists of a circular toroid with an
enhancement in a 3-dimensional blob trailing the Earth.  This
representation ignores the fact that the trailing blob
follows the Earth in an equally eccentric orbit.  Neglect of this
effect is expected to introduce only a small error, though it 
will affect a large range in ecliptic latitude.  This effect may be
worth consideration in future models.  As with the smooth cloud and
dust bands, the symmetry plane of the ring complex (ring + trailing
blob) was allowed to be inclined with respect to the ecliptic plane;
the vertical distance above the ring midplane is denoted by $Z_{R}$,
and is computed using equation~5, with the optimized values of
$\Omega_{RB}$ and $I_{RB}$. The three-dimensional ring dust density
distribution is modelled as 
\begin{eqnarray}
n_R(X,Y,Z)&=&n_{SR}\exp\left[-{(R-R_{SR})^2\over2\sigma_{rSR}^2}
                           -{|Z_{R}|\over\sigma_{zSR}}\right]\\ 
          &+&n_{TB}\exp\left[-{(R-R_{TB})^2\over2\sigma_{rTB}^2}
                           -{|Z_{R}|\over\sigma_{zTB}}
           -{(\theta-\theta_{TB})^2\over2\sigma_{\theta TB}^2}\right]
        \nonumber         
\label{eqring}
\end{eqnarray}
where the subscript ``SR'' stand for the circumsolar ring and ``TB'' stands
for the trailing blob.  The $\sigma$ values are free parameters for
scale lengths in the $R$, $Z_{R}$ and $\theta$ coordinates.  Also free
are the radial locations of the peak density of the ring ($R_{SR}$)
and blob ($R_{TB}$) and the peak densities $n_{SR}$ and $n_{TB}$.

The angle $\theta$ is the heliocentric ecliptic longitude relative to
the mean longitude of the Earth.  We have presumed that the ring
structure is fixed with respect to the mean longitude of the Earth and
not its true longitude.  In these coordinates, the Earth moves in an
epicycle about its mean longitude with an amplitude of $2^\circ$.  The
location of the trailing enhancement is $\theta_{TB} \sim 10^\circ$,
so the epicyclic motion of the Earth changes its distance from the
enhancement by $\sim10-20\%$ over the year.  Isodensity contours of
the dust ring model are shown in Figure~4d.  The map in Figure~4d cuts
through the ring at 1 AU in two places, yielding the two
cross-sectional slices.  Note that the orientation of the map is
rotated about the $Z$ axis by $\theta_{TB}$ so that it cuts through
the trailing blob.

\subsection{Fitting Technique}

The parameters of the IPD model are determined using the
Levenberg-Marquardt non-linear least-squares optimization algorithm
(Bevington 1969).  The fit is constrained using only the observed time
variation along independent lines-of-sight, while ignoring their
underlying photometric baselines.  This is achieved by forcing the
mean of the model over all time samples to match the mean of the data
for each individual line-of-sight.  In this way, the optimization
procedure only has enough information to match the amplitude and phase
of the temporal variation for each line-of-sight, with no assumption
about the morphology or spectrum of non-varying components.  The
goodness of fit is defined so that the model is optimized to match the
modulation as follows. Let $t$ be the observation time (an index over
the weekly sky maps), $p$ be the celestial position (an index over
selected pixels), and $\lambda$ be the wavelength (an index over
spectral bands).  The observed brightness is $I_{\lambda}(p,t)$, and
the model evaluated for the same conditions is $Z_{\lambda}(p,t)$.
Then the goodness of fit is defined as
\begin{equation}
\chi^2=\sum_{\lambda,p,t}{1\over\sigma_\lambda^2(p,t)}
\bigl\{[I_\lambda(p,t)-\langle I_\lambda(p,t)\rangle_t]-
[Z_\lambda(p,t)-\langle Z_\lambda(p,t)\rangle_t]\bigr\}^2.
\end{equation} 
In this equation, $\sigma_\lambda^2(p,t)$ is the estimated
uncertainty on $I_{\lambda}(p,t)$, as derived from the quadrature sum
of random measurement errors and the uncertainty in the temporal
stability of the gain calibration.  A typical value for $\sigma_\lambda$
in the 1.25 -- 4.9 $\mu$m bands is .005 MJy/sr.  The 12 -- 100 $\mu$m
sigma values are dominated by the temporal gain calibration uncertainty:
typical values
are best expressed in terms of ${\sigma_\lambda}\over{I_\lambda}$, and
are $\sim$.0076, .0076, .015 and .022, respectively.  The 140 and 240
$\mu$m bands, which are dominated by instrumental noise, have typical
sigmas of 7 and 4 MJy/sr.

The parameters $T_0$, $A_{c,\lambda}$ and $E_{c,\lambda}$ are highly
correlated with each other, and cannot be determined independently.
The value of $T_0$ was determined by running a preliminary fit which
assumed that the smooth cloud is the dominant component and that the
spectrum in the mid-IR is that of a pure blackbody; i.e., the
albedos and emissivities at 4.9, 12 and 25~$\mu$m were set to
$A_{\lambda} = 0$ and $E_{\lambda} = 1.0$, and the model 
solved for $T_0$.  This value of $T_0$ was then fixed
and subsequently used in all further fits for geometry and source
terms.

Initially, the complete set of model parameters (geometry
and source terms) were determined simultaneously using data from all
10 DIRBE wavelength bands.  In order to avoid excessive computational
requirements, the fitting dataset was chosen from a subset of the 41
weekly-averaged DIRBE maps.  A spatial grid was established for each
wavelength which sampled a sky pixel every $\sim5^\circ$ for ecliptic
latitudes $\leq30^\circ$, and every $\sim10^\circ$ above that.  In
addition to excluding observations within $10^\circ$ of the Galactic
plane, only ``quiet'' pixels which were not overtly situated on a
strong photometric gradient were chosen for each wavelength.  All
available high quality weekly averages were used for each chosen
line-of-sight; this translates to a maximum of 40 time samples (for
pixels near the ecliptic poles) and an average of about 20 time
samples for pixels near the ecliptic plane.  No chosen line-of-sight
had fewer than 8 time samples.  Each wavelength used $\sim 800$ lines
of sight after exclusion of the Galactic plane.
The ``quiet'' pixel
criterion did not constrain the $\sim 800$ lines of
sight to be precisely the same at each wavelength, since
local photometric gradients are also wavelength dependent.

Further analysis showed that the relatively high detector noise and
small contribution of the IPD signal at 140 and 240~$\mu$m caused
these bands to possess very little influence on the derived model
geometry.  In addition, the number of longer wavelength data points
used in the fitting dataset was insufficient to overcome inherent
random errors in these data.  Ultimately, these data constraints
forced adoption of a two-step fitting procedure.  In the first step,
the complete set of model parameters (geometry and source terms) were
found simultaneously using data from the $1.25-100~\mu$m wavelength
bands.  In the second step, the model geometry found
for $1.25-100~\mu$m was assumed to be applicable to the 
140 and 240~$\mu$m bands. 
In this case, the non-linear problem reduces to a linear least-squares
solution for the 140 and 240~$\mu$m model source terms.  
Rather than sampling the sky every few degrees, the complete DIRBE weekly 
dataset at these two wavelengths is used in a separate fit to derive the 
140 and 240 $\mu$m source terms, once again relying only upon
temporal variations.

\section{RESULTS}

The result of the modelling process described above is a formulation
which can be used to describe the modulations in the IPD foreground
observed by DIRBE to an accuracy of 2\% or better.  This section
describes the results of the fitting process, starting with the
small subset of data used to produce the model, and working up to the
production of maps of the full sky in which the zodiacal foreground
has been removed based on our model.  Consistency of the model with
the observed modulation and observed structure in the IPD is also
discussed in this section, whereas the question of the uniqueness of
the model is addressed in \S~6.

\subsection{Optimized Model}

There are nearly 90 potential free parameters in this model, including
both density terms and particle thermal emissivity and scattering
properties.  In the final optimization run, fewer than 50 of the
parameters were actually determined from the fitting procedure.
Treatment of the albedos and emissivities has been been discussed in
\S~4; in addition, some of the dust band and circumsolar ring density
shape parameters were held fixed.  This was necessitated in part by
limitations in the fitting dataset, and in part because of
deficiencies in the analytical model which would cause one density
component to attempt to compensate for inadequacies in another.

The idea of using only the apparent time variation to determine the
zodiacal light was tested on simulated DIRBE observations (including
noise) of a model IPD cloud.  The fitting procedure was able to
recover the properties of the smooth cloud and circumsolar ring
well. However, some angular variations of the low-contrast dust bands
were difficult to recover, partly because of the large grid spacing in
the fitting dataset.  To remedy this, some of these parameters were
set to values derived by Reach \etal (1997) using angular filtering
techniques to isolate the dust band structure.  In addition, tests on
real data showed that the circumsolar ring component would sometimes
be fit in a way that attempted to flatten it to compensate for the
dust bands.  For this reason, the parameters $\theta_{TB}$ and
$\sigma_{rSR}$ were fixed to values derived from visual examination of
DIRBE data.

The final $1.25-100~\mu$m fitting dataset consisted of $\sim 800$
lines of sight per wavelength and a total number of 87,035
observations (including all wavelengths and weekly samples).  This
constitutes only $\sim 0.13$\% of the weekly averaged dataset.  There
are 45 model parameters, leaving 80,425 degrees of freedom for the
model fitting.  Final parameter values are summarized in Tables~1 and
2, together with a short description of each parameter.  Also listed
is the 68\% joint confidence uncertainty for each fitted parameter,
which is derived from the maximum projection of the principal axis of
the 45-parameter 68\% confidence ellipsoid onto the axis for that
individual parameter.

The final total $\chi^2$ per degree of freedom for the optimized fit
to the $1.25-100~\mu$m data was 2.1.  The $\chi^2$ per degree of
freedom achieved for the individual bands from 1.25 to 100~$\mu$m was
2.2, 1.9, 2.0, 3.6, 2.3, 1.5, 1.4 and 1.7, respectively.  These 
reduced $\chi^2$'s are higher than one would expect {\it a priori}.  
Some part of this excess appears to be due to the fact that the brightness of 
the IPD is intrinsically time dependent (see \S~5.6), but this 
variation was not recognized until after the present modelling effort was 
completed.   As discussed in \S~5.6, the intrinsic time variation is small, 
and ignoring it in the model is not believed to have caused significant error 
in the residual sky brightness or the conclusions of our CIB search.

Though the formulation of the IPD modelling problem has been approached 
differently by various researchers, it is interesting to compare the findings 
for some of the major parameters.  This is done in Table~3, where 
comparisons are made with five recent investigations.

\subsection{Residuals within the Fitting Dataset}

Figure~5 shows the IPD model brightness overlaid on the data for three
pixels selected from the fitting dataset used to optimize the model
parameters.  The zodicacal model intensity has been offset to match
the mean observed pixel brightness, just as is done in order to fit
the data.  Within the error bars, the fit looks good on a
pixel-by-pixel basis.  Section 3 and Figure~3 provide an explanation
of the shape of the intensity variation.

The character of the mean-adjusted residuals $[(I_\lambda-\langle
I_\lambda\rangle)-(Z_\lambda-\langle Z_\lambda\rangle)]$ for the
complete dataset used to fit the time variation is illustrated in
Figure~6.  The histograms in Figures~6a-d show that the overall means of
the residuals from the fit are very close to zero within a roughly
Gaussian distribution function of narrow width.  A search for
systematic trends in the residuals versus ecliptic
latitude and solar elongation is presented for 1.25, 3.5, 25 and 100~$\mu$m
in Figure~6.  The 25~$\mu$m trend with solar elongation shows some 
systematic structure, which is also seen at 12~$\mu$m.

\subsection{Residuals for a Weekly Map}

The optimized DIRBE IPD model brightness was evaluated for each of the
41 Weekly Sky Maps described in \S~2 using the model parameters
(Tables~1 and 2) and equation~1.  Figure~7 illustrates the IPD
modelling results for one of these weekly maps (DIRBE mission week 22,
which consists of averaged observations over the interval 1990 April
16-22, inclusive).  For each pixel, the model was evaluated for the
same time as the average time of observation of that pixel during the
week.  The results for the 25~$\mu$m data are shown in order to
illustrate most clearly deficiencies of the model.  The observed sky
brightness is presented in Figure~7a, and the corresponding IPD model
intensity for this week is shown in Figure~7b on the same scale.  The
residual map which results from subtracting the IPD model from the
data is provided in Figure~7c on a linear scale whose maximum is $\sim
20$ times fainter than that used to display the observed sky
brightness.  The brightest residual features are emission from the
Galactic plane.  There are, however, lower-level systematic residuals
which arise from imperfections in the zodiacal model.  The most
noticable of these residual imperfections lie near the ecliptic plane
and solar elongation extrema.  An excess residual in the 12 and 25~$\mu$m 
maps at solar elongation extrema was also indicated in the
results for the fitting dataset described in \S~5.2.

Individual contributions to the IPD model brightness from the smooth
cloud, dust bands, and circumsolar ring components are given in
Figures~7d, 7e, and 7f, respectively.  The intensity stretch for
Figures~7e and 7f is expanded $\sim 20$ times compared to Figure~7d in
order to show detail.  The intensity from the circumsolar ring is
asymmetric with respect to the Earth-leading and trailing halves of
the sky; the brighter half corresponds to that containing the trailing
``blob'' (\S~4.2.3).  Comparison of the residuals in Figure~7c with
the individual component contributions in Figures~7d and 7f shows that
some of the lower-level structure in the residuals correlates with the
dust bands and circumsolar ring intensity contours.  There is an
ecliptic plane excess in the leading half of the week 22 residual map
near elongation $80^\circ$.  The similarity of this excess to the
brightness profile of the ring indicates that there is an insufficient
model contribution from the circumsolar ring.

\subsection{Average Residuals}

Mission-averaged maps of the sky with zodiacal light removed were
formed by averaging together the available weekly residual maps, of
which Figure~7c is one example.  Because of the imperfections in the
IPD model at extreme solar elongation angles described previously, a
solar elongation constraint was implemented by which weekly-averaged
observations for which $e\lt68$\deg\ and $e\gt120$\deg\ were excluded
from the average.  In addition, data from three of the 41 weeks were
excluded from the average for the $1.25-100~\mu$m data because of low
numbers of normal survey observations in those weeks ({\it COBE} DIRBE
Explanatory Supplement, 1997).

Figure~2 shows the observed mission-averaged sky brightness at all 10
DIRBE wavelengths, in addition to the mission-averaged maps after
zodiacal light removal.  Mission-averaged maps which have had the IPD
signal removed are shown on two scales --- the first on the same
logarithmic scale as the measured data, in order to illustrate the
success of zodiacal light removal compared to the original signal
(Fig.~2b).  The second scale is a linear stretch chosen to emphasize
any visible defects in low-intensity portions of the residual maps
(Fig.~2c).

If the removal of the IPD signal were perfect, the mission-averaged
residual maps would appear as clean images of the Galaxy and any
extragalactic light.  While the removal of the zodiacal signal in the
near- and far-infrared looks relatively free of defects even on the
expanded linear scale of Figure 2c, there are clearly systematic
defects present in the $4.9-60~\mu$m residual maps.  These wavelengths
present the most rigourous test of the IPD model, since it is at these
wavelengths that the zodiacal foreground is strongest.  In general,
there are systematic residual features parallel to the ecliptic plane
and within about 15\deg\ of it at these wavelengths.  Such features
presumably arise from deficiencies in the model.  In general, these
features are relatively faint.  For example, the 12~$\mu$m excess near
the ecliptic plane is of order 2\% of the zodiacal signal.

\subsection{Contribution of IPD Signal to Total Sky Brightness}

The IPD model provides quantitative estimates of the contribution of
the zodiacal foreground to the observed sky brightness at each of the
ten DIRBE wavelengths.  Figures~8 and 9 illustrate the observed sky
intensity and corresponding IPD model intensity for representative
directions in the sky.  For regions not dominated by strong Galactic
sources, the estimated fraction of the total sky brightness due to
zodiacal light varies from about 2/3 in the near-infrared ($1.25 -
3.5~\mu$m) to more than 90\% in the mid-infrared ($4.9 - 60~\mu$m).
Even in the far-infrared, the zodiacal contribution is not completely
negligible at high Galactic latitudes: $\sim 20$\% of the 240~$\mu$m
emission in the ``Lockman hole'' (direction of minimum HI column
density, Lockman \etal 1986) can be attributed to zodiacal emission,
as shown in Figure~9b.  The values used to make the plots in Figure~9
are also presented in Table~4, with further details about the
locations in the figure caption.

Quantitative uncertainties associated with the IPD model are discussed
in the following sections.  These include both the errors associated
with fitting the observed time dependence with the chosen model,
as well as uncertainty in the true IPD brightness resulting
from the fact that other model formulations might fit the time
dependence of the observations equally well.  This issue of model
``non-uniqueness'' is of critical importance for cosmological studies
such as the search for the CIB.  As a simple illustration, the
zodiacal brightness at high ecliptic latitudes is influenced at the
few-percent level by the modelled contribution of the near-Earth IPD
structures.  Changing the formulation for the 10\deg\ asteroidal bands
so that they migrate in completely toward the Sun, rather than having
an inner cut-off radius as in our model, increases the contribution
for the dust bands shown in Figure~9b by a factor of 5 at high
ecliptic latitude.  A change in the band inner cut-off has little
effect on the model at low ecliptic latitudes.  Changes in the
residual sky brightness resulting from different inner radial cut-offs
for the bands are within the uncertainties of the isotropic components
of the IPD model discussed in \S~6.

\subsection{Consistency of Model as a Function of Time}

It is difficult to specify precisely the errors inherent in our
modelling technique of fitting the time variations of the sky
brightness.  This is particularly true because the level and nature of
the errors depend both on epoch and spatial location.  To give some
insight on this point, Figure~10 presents maps of the 12~$\mu$m sky brightness
in ecliptic coordinates for four weeks during the mission.  These
weeks have been selected so that two of the maps were obtained
approximately one-half year after one of the earlier maps.  Data at 
12~$\mu$m were used because the IPD signal is the predominant
signal in the sky and because all features in the model are evident in
the data.  These specific images are ``constant-time maps'' created by
interpolating linearly between contiguous weekly-averaged maps (in
which the mean time of observation of each pixel is not a constant,
but depends upon the times in which it was scanned during that week).  
In ``constant-time'' maps the brightness at each pixel represents a view of 
the IPD cloud at a fixed time, thus eliminating the small brightness 
fluctuations in our Weekly Sky Maps arising from different mean 
observation times per pixel.

These constant-time maps have been used to create Figure~11, which
shows the variation of the brightness of the sky, the IPD model, and
the residuals (difference between measured brightness and model) as a
function of ecliptic longitude for two representative latitudes.
This figure also shows the average of the weekly residual results.
Both the sky maps and line plots disclose much about the nature of the
modelling errors.

Comparing the maps separated by half a year shows that the residuals
in the ecliptic polar regions are quite comparable, but there are
clear indications of variable errors in the vicinity of the ecliptic
plane.  Note in particular the large-scale inversions of the strength
of the residuals between maps separated by half a year in the center
portion of the maps.  This results mainly from an inadequate
representation of the circumsolar ring and the Earth-trailing blob.  
Errors resulting from
deficiencies in the model of the dust bands are also visible.  While
the errors can be easily seen in these maps, it is essential to
realize that these obvious artifacts result from modelling errors
which are only a few percent of the intensity of the zodiacal
emissions.  But the nature of the errors is complex; some features of
this complexity are illuminated by the latitude-cut line plots of
Figure~11.

In the plots for $\beta=80^\circ$, it is seen that the longitudinal
dependence of the zodiacal emission is seemingly very well tracked by
the model.  But there is a noticeable phase shift between the
observations and the model.  This phase difference is vividly
translated into a wide separation of residuals over a half-year
interval.  Though not shown graphically, for weeks 6 and 32 
this separation reaches a
maximum around longitude 180\deg, whereas for weeks 12 and 
38 the peak separation appears around longitude 240\deg.  
The cause of this temporal variation in the phase
shift between the observations and the model is not known.

The model does reproduce well the striking differences in the
amplitude of the time variation of the IPD signal in different
directions.  For example, in the data of weeks 6 and 32, the variation
in the sky at a latitude of 80\deg\ for longitude of 100\deg\ is
$\sim0.1$~MJy~sr$^{-1}$.  This is in contrast to the observations at
270\deg\, where the variation is $\sim3.7$~MJy~sr$^{-1}$.  The range
of temporal variation at high latitudes is strongly dependent on
longitude.  That the model captures this distinctive feature of the
temporal variation indicates that the basic geometry of the IPD cloud
is represented reasonably well in the model.  A comparison of the
weekly results with the mission-averaged values indicates that the
error in the modelled zodiacal contribution ranges from 0 to 4\% of 
the IPD modelled brightness, implying an error in the average 
residual ranging from about 0 to 42\%.

The plots for $\beta=0^\circ$ demonstrate what can be seen in the
corresponding full-sky maps --- there are strong anti-correlated
variations in the residuals separated by one-half year.  All elements
of the model, i.e., main cloud, circumsolar ring, and bands, combine
to produce these effects.  Though the residuals behave quite
differently in the two halves of the year, it should be noted that the
differences in the residuals only represent a modest lack of fidelity
in the model, ranging from about 0 to 3.8\% of the IPD brightness in
a pixel.

The characteristics of the DIRBE data set which make it well-suited
for modelling of the IPD signal based upon the observed time variation
of the sky brightness (highly repetitive sampling, stable photometry)
also permit some unique analyses of the global deficiencies in the
modelling.  One of the best methods for discerning the general nature
of the model defects over the whole sky comes from an analysis of the
weekly residuals maps.  As noted above, a grand residual sky map was
produced at each wavelength by subtracting the weekly map calculated
from the IPD model from the map of weekly-averaged observations, and
then averaging these weekly residual maps over all the weeks of the
mission as described in \S~5.4.  
Using the average residual map as a fiducial measure,
a distribution function representing the quality of each weekly residual 
map was constructed by forming the difference between the  weekly
residual map and the mission-averaged residual map, expressed as a fraction of
the IPD model brightness for that week:
\begin{equation}
\{[I_\lambda(w,p)-Z_\lambda(w,p)]-\langle{I_\lambda(w,p)-Z_\lambda(w,p)}
\rangle_w\}/Z_\lambda(w,p),
\end{equation}
where this expression is evaluated for each visible pixel $p$ in the
weekly map.  If the modelling were perfect, the distribution functions
of these weekly pixel differences would be Gaussians with mean zero.

Using again the 12~$\mu$m data for illustration, Figure 12 shows two
examples of such distribution functions for mission weeks 12 and 38.
These distribution functions are created after removing the bottom and
top 0.5\% of the data to remove the effects of discrepant outliers.
The number of pixels contributing to such distribution functions is a
bit more than those in half the sky, i.e., about 205,000 per week.  As
can be seen in Figure 12, the distribution for week 12 is not Gaussian whereas 
the distibution for week 38 is nearly Gaussian.  These two examples also 
illustrate that the means of the
distribution functions are not stable with time.  This instability in
the means is shown in Figure 13, where the time variation of the means
of the weekly distribution functions are shown for the 1.25, 3.5,
12, and 240~$\mu$m data.  A five-term harmonic fit to the means with a
one-year period is also shown.  The use of a one-year fitting period is 
based on the conclusion that one problem with the model arises from assuming
the circumsolar ring to be circular, whereas the orbit of the Earth about the Sun 
is eccentric.  It is interesting to note that the
phase of the fit for the 12~$\mu$m data is almost 180 degrees out of
phase with those for the other wavelengths. It is believed that this
could be caused by the strong dependence of the 12~$\mu$m result on
the assumed model structure for the circumsolar ring and trailing blob.

The plots in Figure 13 also show a high frequency variation of the means, 
with a period of the order of 27 days and an amplitude of $1-2\%$.  Such
variations are not seen in the photometry of the discrete standard
objects used to stabilize our photometric system on long time scales,
so they are not artifacts arising from the calibration process.  Such
variations are also seen in the average behavior of the deviations of
subsets of isolated pixels relative to harmonic fits of the
observations, i.e., it is not a phenomenon arising only in some
special direction(s) in the sky or from the modelling of the IPD cloud.  
It is most strongly seen in the 
4.9~$\mu$m data.  The approximate 27 day period suggests a cause related
to the Moon or to solar rotation.  Since the Moon crosses the DIRBE
scan swath twice per lunar month, any lunar effect would be expected to
have a 14 day period.  Furthermore, the DIRBE off-axis response was 
measured in flight and found to be orders of magnitude less
at all angular distances from the Moon retained in the DIRBE data
analysis than the residual effects seen in Figure 13.  A more likely
cause is therefore solar modulation of the zodiacal light brightness due to
solar rotation and variation of the UV flux associated with sunspots.  
This is supported by the fact that the variations are correlated with 
the Mg~II absorption strength.  A similar UV-flux variation periodicity 
is seen in the sky Ly-$\alpha$ background (Qu\'{e}merais, Sandel \&
de Toma 1996).  A more thorough discussion of this discovery 
will be presented in a separate paper (Kelsall \etal 1998).  Although low- and
high-frequency variations are evident in the residuals, these effects
are small, and the modelling results at all wavelengths reproduce the
mean behaviour of the observed sky to within a few percent of the
level of the IPD contribution.  Table~5 summarizes the deviations of the 
weekly residual sky maps from the mission-averaged residual sky maps, as 
illustrated in Figure 12.  Table~5 also shows similar statistics for the high
quality region B used in our search for the CIB (Paper III).

\section{Uncertainties in Cosmological Results}

The uncertainties and errors in the model of the IPD thermal emission
and scattered light can be characterized in many different ways;
e.g., measurements of residual variations as a function of time, or
amount of residual emission that is correlated with components of the
IPD model.  The errors evident as temporal variations of the residual
emission are reduced by averaging the data over time for the length of
the DIRBE cold-era mission.  Model errors that are apparent as
residual structure in the maps can often be minimized by selection of
``good'' regions of the maps (typically at high latitude) for use in
further analyses. However, since we ultimately seek to determine the
level of an isotropic CIB, the most significant form of uncertainty is
that which affects the amount of emission from any potential isotropic or
nearly-isotropic IPD component.  This is an issue of model uniqueness:
Are there any other models that can give an equally good fit to the
apparent time variations of the IPD emission and scattered light, and 
yet lead to substantially different isotropic residuals?

To evaluate the uncertainty of a CIB measurement due to the
uncertainty in the isotropic or nearly-isotropic components of the IPD
model, we attempted to fit the IPD cloud using several different
functional forms (kernels) for the density distribution of the main IPD 
cloud.  The intensity
differences between models which fit the time variations equally well
are taken as a measure of the uncertainty in the IPD model.

For this analysis, the shape and position of the small scale
components of the IPD were fixed, except for the inclination and line
of nodes of the $10^\circ$ dust bands.  Functional forms for the main
cloud that were tested included the {\it IRAS} model (Wheelock \etal
1994), a modified fan model, an ellipsoid model, a sombrero model
(Giese \etal 1986) and a widened modified fan model (eqs. 6 and 7).
Each of the models was used to fit the time variations of the $1.25 -
100~\mu$m intensities at one pixel in each $5^\circ\times5^\circ$
patch of the sky.  Regions within $10^\circ$ of the Galactic plane
were excluded.  In other respects the fitting of each model was
performed in a manner similar to that described in \S~4.3.
In general the differences
between the models are encouragingly small, being of the
order of a few percent.  This is illustrated in Table~6,
where the relative values of the sky brightness as a 
function of model is shown for the north Galactic pole.

For the sample of functional forms described above, the total $\chi^2$ per 
degree of freedom across all
fitted wavelengths was comparable for the modified fan ($\chi^2 =
2.292$), the widened modified fan ($\chi^2 = 2.273$), and the
ellipsoid ($\chi^2 = 2.253$) models.  The {\it IRAS} and sombrero
models fared distinctly worse at $\chi^2=2.562$ and 3.187,
respectively.  The model intensities were evaluated at five
high-latitude regions of interest for measurement of the CIB
(Paper~I): the north and south ecliptic poles, the north and south
Galactic poles, and the Lockman hole region, which contains the
location of lowest Galactic H I column density.  Among the three
models with the lowest $\chi^2$, the largest intensity differences
were found at the NGP region.  At wavelengths from $1.25-100~\mu$m the
ellipsoid model was brighter than the widened modified fan model,
which was brighter than the modified fan model.  Thus, we have taken
the differences between the ellipsoid and the modified fan models at
the NGP as an indication of the uncertainty in the IPD models.  These
uncertainties are listed in Table~7.  The uncertainties at 140 and 240
$\micron$ were obtained by scaling the 100 $\micron$ uncertainty by
the mean color of the IPD emission at 100 $\mu$m relative to the emission
at 140 and 240 $\mu$m, respectively.

By choosing the {\it largest} variations among good IPD models at high
latitudes, we may be overestimating the uncertainty of the IPD model
at a {\it typical} high latitude location.  However, we cannot be
certain that there are not other models of the IPD cloud that can fit the
time variation of the data as well as the models we have investigated, and 
yet lead to larger differences in the predicted model intensity.  Most 
important, the existence of a
real isotropic component in the IPD contribution to the sky brightness
which is not represented in any of these semi-empirical models cannot
be discerned directly by fitting the time variation in the DIRBE data.
Independent arguments limiting the likely intensity of such a
component are made in Paper~IV and summarized in Paper~I.

An additional systematic error that needs to be recognized arises from
the fact that, as noted in \S~5.6, the weekly mean deviations of the
residual sky maps (after the modelled IPD contribution is removed)
from the mission-averaged residual map show a strong harmonic feature
with a period of 1 year.  Since the DIRBE observations did not cover a
full year, averaging this harmonic variation over an incomplete cycle
introduces a small systematic ``truncation'' error.  
These errors have been computed for both the whole sky and a smaller region
used for the analysis in Paper I.  Since the main cosmological results
of Paper I are based upon small selected areas of the sky, it is important
to check whether the magnitude of the error seen for the whole sky applies
to these selected areas.  A study of the distribution functions of the 
deviations of the weekly residuals
was performed for the ``High Quality Region B (HQB)'', which contains
two patches at high northern and southern Galactic and ecliptic
latitudes.  The variation of the means of these distribution functions
is similar to that for the whole sky.  The grand average of these
weekly means has a truncation error of about one percent or less
of the IPD contribution, except at 240~$\mu$m, as shown in Table~8.
The truncation error has a generally small impact on the overall
uncertainties of the IPD model at every DIRBE wavelength.

\section{CONCLUSIONS}

This paper describes in some detail the construction of a physically 
motivated, parametric model of the IPD cloud contribution to the 
infrared sky brightness using the most distinctive and only unique 
feature of that contribution: its apparent temporal variation in fixed 
celestial directions induced by the orbiting of the Earth
about the Sun.  The effort to arrive at a model satisfactory for our
search for the cosmic infrared background (CIB) turned out to be more 
arduous than expected, since the data revealed the complex cloud 
structures not known when the DIRBE investigation was planned.  
The model presented here is the survivor from a large number of
parameterizations and fitting procedures which we have 
investigated.  This extensive effort, 
while time consuming, was essential since it was driven by the desire 
to achieve an $\sim1\%$ precision in the identification of the IPD 
signal.  It is critical to note that ``precision'' is the correct word, for 
though the model well represents the time variations, 
it is not a unique model.  It is, in fact, impossible to determine a 
unique model from any set of data taken from within the IPD cloud. 
Only a mission flying well beyond the orbit of Jupiter could gather data
permitting a unique solution.  

Though not unique, the model described here is of
rather high fidelity, in that it reproduces the apparent time variations 
over the sky and leaves quite stable residual maps at all wavelengths.  
The range of functional forms investigated allows an estimate
of the uncertainties due to the lack of uniqueness of the final model.  
The modest size of these uncertainties has allowed successful conduct of
our search for the CIB (Paper I).  The resulting residual maps with the IPD
contribution removed are the best all-sky absolutely-calibrated photometric 
maps now available 
for study of large-scale Galactic phenomena from 1.25 to 240 $\mu$m, though
evident artifacts do remain, especially at low ecliptic latitude.

The virtues of our approach for finding the contribution of the IPD cloud 
to the infrared sky brightness include the following:

\begin{itemize}
\item Modelling the time variations only is a unique method which does
not require extraneous zero point constants, and automatically excludes 
contributions from Galactic or extragalactic sources.

\item Modelling using the data from many wavelengths simultaneously
provides an IPD structure which is consistent with the data and the same for 
all wavelengths.

\item The IPD cloud density function kernel found by modelling the
time variations produces a level for the IPD contributions over two
orders of magnitude in wavelength which are compatible with
expectations, i.e., that a positive sky residual remains after
removing the modelled IPD contribution.

\item The isotropic part of the model is only modestly sensitive to
the functional forms chosen to parameterize the model.

\item The method readily permits incorporation of all known elements
of the IPD cloud: a main cloud, the asteroidal bands, and a
circumsolar ring containing dust particles in Earth-resonant orbits
near 1 A.U.

\item The model yields a global precision of the order of a few
percent of the IPD contribution. 

\item Because the IPD cloud contains reasonably smooth features, the method
is efficient in that a robust IPD model can be constructed using only a
very limited fraction ($\sim0.13\%$) of the DIRBE data set.
\end{itemize}

Though our approach produced a successful model for the IPD cloud, it
is also clear that it is not a perfectly consistent model for the
whole sky.  Testing the time stability of the weekly residual sky maps
after subtracting the modelled IPD contribution is a powerful tool for
assessing the quality of the model.  The result of that assessment is 
that the model did achieve the goal of matching the time variations.  
There are, however, apparent spatial artifacts in the results that are
correlated with the various components of the IPD cloud and quite
clearly reflect imperfections in the model.  On the basis of this
work, we feel that further improvements could be made in constructing
a quasi-physical model that, without recourse to elaborate
representations of the dust, should be able to achieve further 
improvements in the results.

Some possible changes in the modelling technique, in rough order of
perceived priority, are as follows: 

\begin{itemize}

\item Force the trailing blob in the circumsolar ring to trail the Earth in its
eccentric orbit;

\item Include a separate contributing form factor for the dust coming
from the main-belt asteroids; 

\item Incorporate more realistic representations for the asteroidal
bands and the circumsolar ring, where now, for example, the
representations use simple Gaussian distributions; 

\item Include a source function scattering term at 4.9 and 12 $\mu$m since a
significant fraction of the IPD particles are large enough to scatter
these wavelengths efficiently;

\item Recognize a possible warpage of the symmetry plane as a
function of distance from the Sun due to the
influences of Jupiter, Mars, Earth, and Venus; 

\item Include an intrinsic  time variation of the IPD brightness 
with a period equal to that seen in the data;

\item Permit variation of the albedo for the shorter wavelength
bands to accommodate the clues in the observations that point
to a variation with latitude, which may well result from the 
differences in the dust contributed by comets as compared to that 
coming from asteroids;

\item Add a contribution from a hot dust component; and,

\item Incorporate the deviations from the grand average residuals
directly into an iterative modelling loop so as to limit the available
solution space and push the results towards distribution functions of 
residuals which are Gaussian. 
\end{itemize}

The data described in this paper are available to the public
from the NSSDC through the COBE website at 
{\tt http://www.gsfc.nasa.gov/astro/cobe/}.  The DIRBE weekly
maps of the sky and the corresponding IPD model brightness are
available as the ``DIRBE Sky and Zodi Atlas (DSZA)'' product; the
mission-averaged, zodi-subtracted residual skymaps are contained
in the ``Zodi-Subtracted Mission Average (ZSMA)'' maps.

\acknowledgments

The authors gratefully acknowledge the contributions over many years
of the many engineers, managers, scientists, analysts, and
programmers engaged in the DIRBE investigation.  The
National Aeronautics and Space Administration/Goddard Space Flight
Center (NASA/GSFC) was responsible for the design, development, and
operation of the {\it COBE}.  Scientific guidance was provided by the
{\it COBE} Science Working Group.  GSFC was also responsible for the
development of the analysis software and for the production of the
mission data sets.

\clearpage

\figcaption[]{DIRBE 12~$\mu$m maps of the sky for various integration
periods: a) one orbit; b) one day; c) one week; d) 10 months. 
Maps are Mollweide projections in ecliptic coordinates.}

\figcaption[]{DIRBE mission-averaged sky brightness in 10 wavebands,
both before and after removal of the IPD signature: a) as-observed
sky, logarithmic scale; b) sky after removal of ZL, scaled identically
with a); c) same as b), but on a linear scale expanded to show defects
in the residuals. Units are in MJy~sr$^{-1}$. Sixteen color contour
levels are used. The minimum and maximum scaling values used for each
band are listed as follows, with the first set of numbers being the
logarithmic scaling limits, and the second set being the linear scale:
1.25~$\mu$m: [.063,31.6], [-0.05,0.3]; 2.2~$\mu$m: [0.04,31.6],
[0,.2]; 3.5~$\mu$m: [0.032,31.6], [-0.01,.2]; 4.9~$\mu$m: [0.1,15.8],
[0,0.2]; 12~$\mu$m: [1.58,79.4], [0,2]; 25~$\mu$m:
[3.98,79.4],[0.5,3]; 60~$\mu$m:[1,79.4], [0,3]; 100~$\mu$m: [1,79.4],
[0,15]; 140 and 240~$\mu$m: [1,794.3], [0,20].
All maps are Mollweide projections in ecliptic coordinates.}

\figcaption[]{Schematic representation of zodiacal light brightness
vs. time as observed by DIRBE at the: a) north ecliptic pole and b)
ecliptic plane.  In a), the dust cloud is represented as a shaded
ellipse which is tilted with respect to the plane of the Earth's
orbit.  (See text for additional explanation.)}

\figcaption[]{Isodensity contours of the IPD Model Components, shown for
a cross-sectional slice perpendicular to the
ecliptic plane: a) all components combined; 
b) smooth cloud; c) dust bands; d) circumsolar
ring. The density contour levels used in a) and b) are listed in brackets
at the bottom of a), in units of $10^{-7}$ AU$^{-1}$.  Contour levels used
for c) and d) are a factor of eight smaller.}

\figcaption[]{Sky brightness vs. time as observed by DIRBE for 4 
different wavebands and three different sky locations: a) the north
ecliptic Pole; b) the north Galactic pole and c) the ecliptic plane
(near ecliptic longitude 180\deg).  Error bars include
random and systematic uncertainties.  Smooth curve through the data is
that generated from the IPD model, with an offset added to raise
the zodiacal contribution to the mean of the observed pixel photometry.}

\figcaption[]{Overall character of residuals in the fitting dataset for
four representative wavelengths:  a) 1.25~$\mu$m, b) 3.5~$\mu$m,
c) 25~$\mu$m and d) 100~$\mu$m.  For each wavelength band, the
histogram of residuals,  residuals as function of ecliptic
latitude and residuals as function of solar elongation are plotted from
bottom to top.}

\figcaption[]{Data, IPD model and residual maps for Week 22 (90106-90112) at
25~$\mu$m:  a) the as-observed sky. The scale is linear: [0,100]
MJy~sr$^{-1}$;  b) sky brightness for this week as predicted by the
IPD model, on the same scale as a).  The IPD model intensities shown
here are broken down into the brightness due to the smooth cloud,
bands, and ring in d), e), and f), respectively;  c) the residual map 
(observed - model), on a linear scale [0,5] MJy~sr$^{-1}$;  d)
Brightness for the smooth cloud model component, on the same scale as
a); e) Brightness for the combined bands components, scale=[0,5.5];
f) Brightness for the ring model component, scale=[0,5.5].
All maps are Mollweide projections in ecliptic coordinates.}

\figcaption[]{IPD contribution to the observed sky brightness.  The
lower curve in each plot is the ZL brightness for the time and
locations computed using the DIRBE IPD model.  a) DIRBE observations of
the infrared sky brightness on 1990 Jan 19 at solar elongation =
$90\deg$, ecliptic longitude $179\deg$.  The 140 and 240~$\mu$m data
have been averaged and smoothed.  b) Mission-averaged intensity profile
in the Galactic plane.}

\figcaption[]{Spectral energy distribution of the observed sky vs. the
IPD model.  The contributions from individual density components of
the zodiacal cloud are also indicated.  a) for DIRBE pixel 162811, located
on the ecliptic plane at ecliptic longitude 122\deg.  The time of
observation is $\sim$day 109 of 1990, corresponding to a solar
elongation of approximately 90\deg.  b) for DIRBE pixel 64552, located
within the Lockman Hole (the region of minimum HI column density):
ecliptic coordinates (137\deg,46\deg); Galactic coordinates
(148\deg,53\deg). The time of observation is $\sim$day 129 of 1990,
corresponding to a solar elongation of approximately 90\deg.
(See Table~4 for the numerical values.)}

\figcaption[]{DIRBE 12~$\mu$m (data-model) residual maps for mission week 
pairs spaced
6 months apart: Weeks 6 and 32 (10a,b) correspond to mid-week observations
times of 28 Dec 1989 and 28 Jun 1990, respectively, and are displayed
on a linear stretch of [.456,3.511] MJy/sr.  Similarly, weeks
12 and 38 (10c,d) correspond to 8 Feb 1990 and 9 Aug 1990, respectively,
and the scale = [.439,3.987] MJy/sr. Maps are Mollweide projections in
ecliptic coordinates.}

\figcaption[]{The 12~$\mu$m intensity profiles as a function of ecliptic
longitude for ecliptic latitudes of 80\deg\ (11a) and 0\deg\ (11b).
For each latitude the top two panels show 
the sky data for two weeks separated by 26 weeks, while the next 
two panels show the corresponding data from the IPD model.  The 
bottom panel shows an overlay of the differences of the sky and 
IPD model for each of the two weeks, as well as the run of the
mission-average residual.}

\figcaption[]{Representative distribution functions of the ratio
$\{[I_\lambda(w,p)-Z_\lambda(w,p)]-\langle{I_\lambda(w,p)-Z_\lambda(w,p)}
\rangle_w\}/Z_\lambda(w,p)$
where $\langle{I_\lambda(w,p)-Z_\lambda(w,p)}\rangle_w$
is the mission averaged value.  (a) shows the result for mission week 12 and
(b) shows the same result for week 38; i.e., one displaced by
1/2-year where the expectation would be for similar results, which
clearly are not obtained.}

\figcaption[]{Samples of the run of the full-sky mean for the 
$\{[I_\lambda(w,p)-Z_\lambda(w,p)]-\langle{I_\lambda(w,p)-Z_\lambda(w,p)}
\rangle_w\}/Z_\lambda(w,p)$  ratio as a 
function of time for the a) 1.25, b) 3.5, c) 12 and d) 240~$\mu$m bands.}

\clearpage





\setcounter{table}{4}

\begin{deluxetable}{lcccc}
\label{tbl:table5}
\tablecaption{Deviations of the Weekly Residual Maps from the 
Mission-Averaged Residual Maps (see equation~11)}
\tablehead{
\colhead{Region}&
\colhead{Band}&
\colhead{$\langle{Mean}\rangle$}&
\colhead{$\langle{|Mean|}\rangle$}&
\colhead{$\langle{Median}\rangle$}\\
\colhead{}&
\colhead{(\um)}&
\colhead{(\%)}&
\colhead{(\%)}&
\colhead{(\%)}
}
\startdata
Full Sky & 1.25 & $ 0.07 \pm 1.76$  & $ 1.49 \pm 1.77$ & $ 0.11 \pm 1.27$ \nl
         & 2.2 &  $0.04 \pm 0.92$   & $ 0.73 \pm 0.92$ & $0.09 \pm 0.60$ \nl
         & 3.5 & $-0.11 \pm 1.59$   & $1.26 \pm 1.59$ & $-0.04 \pm 1.09$ \nl
         & 4.9 & $ 0.35 \pm 1.07$   & $ 0.88 \pm 1.12$ & $ 0.17 \pm 0.92$ \nl
         & 12  & $ 0.30 \pm 0.68$  & $ 0.57 \pm 0.74$ & $ 0.17 \pm 0.71$ \nl
         & 25  & $ 0.20 \pm 0.30$  & $  0.28 \pm 0.36$ & $ 0.11 \pm 0.30$ \nl
         & 60  & $ 0.10 \pm 1.25$  & $ 0.99 \pm 1.26$ & $ 0.07 \pm 1.05$ \nl
         & 100 & $ 0.52 \pm 4.10$  & $  2.74 \pm 4.14$ & $ 0.03 \pm 2.45$ \nl
         & 140 & $ 0.82 \pm 12.85$ & $  9.00 \pm 12.88$ & $ 0.24 \pm 10.11$ \nl
         & 240 & $ 0.77 \pm 21.22$ & $ 16.6 \pm 21.23$ & $ 1.23 \pm 16.38$ \nl
\tableline
High     & 1.25&  $ 1.34 \pm 2.63$  & $ 1.41 \pm 2.07$ & $ 0.99 \pm 2.54$ \nl
Quality  &  2.2 & $ 1.04 \pm 1.44$  & $ 0.53 \pm 0.91$ & $ 0.70 \pm 1.47$ \nl
Region B &  3.5&  $ 1.64 \pm  1.96$ & $ 0.88 \pm 1.28$ & $ 0.46 \pm 2.39$ \nl
         &  4.9&  $ 1.60 \pm  1.70$ & $ 0.83 \pm 1.37$ & $ 0.75 \pm 2.21$ \nl
         &   12&  $ 0.84 \pm 1.07 $ & $ 0.37 \pm 0.59$ & $ 0.68 \pm 1.21$ \nl
         &   25&  $ 0.65 \pm 0.82 $ & $ 0.42 \pm 0.55$ & $ 0.53 \pm 0.94$ \nl
         &60   &  $ 0.75 \pm 1.54 $ & $ 0.75 \pm 0.98$ & $ 0.38 \pm 1.68$ \nl
         &100  &  $ 1.51 \pm 2.87 $ & $ 1.07 \pm 1.68$ & $ 0.49 \pm 3.16$ \nl
         &140  &  $ 4.90 \pm 13.51$ & $ 6.29 \pm 8.16$ & $-0.54 \pm 18.51$ \nl
         &240  &  $13.02 \pm 31.31$ & $11.38 \pm 15.70$& $-0.22 \pm 40.55$ \nl
\enddata
\end{deluxetable}

\clearpage

\begin{deluxetable}{ccccc}
\tablecaption{ Relative IPD Cloud Brightness at the North Galactic Pole
               as a Function of Model Type}
\label{tbl:table6}
\tablehead{
\colhead{Wavelength}&
\colhead{Present}&
\colhead{Modified}&
\colhead{Widened Modified}&
\colhead{Ellipsoidal}\\
\colhead{($\micron$)}&
\colhead{Model}&
\colhead{Fan Model}&
\colhead{Fan Model}&
\colhead{Model}
}
\startdata
  1.25 &     1.00  &      1.05   &     1.06 &       1.08 \nl
   2.2 &     1.00  &      1.11   &     1.12 &       1.14 \nl
   3.5 &     1.00  &      1.00   &     1.00 &       1.03 \nl
   4.9 &     1.00  &      1.01   &     1.01 &       1.03 \nl
  12.  &     1.00  &      0.98   &     0.98 &       1.00 \nl
  25.  &     1.00  &      0.97   &     0.98 &       1.01 \nl
  60.  &     1.00  &      0.94   &     0.95 &       0.99 \nl
 100.  &     1.00  &      0.93   &     0.95 &       1.00 \nl
\enddata
\end{deluxetable}

\clearpage

\begin{deluxetable}{cc}
\tablewidth{4in}
\tablecaption{IPD Model Uncertainties for the CIB}
\label{tbl:table7}
\tablehead{
\colhead{Wavelength}&
\colhead{Uncertainty}\\
\colhead{($\micron$)}&
\colhead{(nW m$^{-2}$ sr$^{-1}$)}
}
\startdata
  1.25 & 15   \nl
  2.2  &  6   \nl
  3.5  & 2.1  \nl
  4.9  & 5.9  \nl
  12   & 138  \nl
  25   & 156  \nl
  60   & 26.7 \nl
  100  & 6.3  \nl
  140  & 2.3  \nl
  240  & 0.5  \nl
\enddata
\end{deluxetable}

\clearpage

\begin{deluxetable}{ccc}
\tablewidth{5in}
\tablecaption{Truncation Error in the High Quality Region B}
\label{tbl:table8} 
\tablehead{
\colhead{Wavelength}&
\colhead{Northern Patch}&
\colhead{Southern Patch}\\
\colhead{(\um)}&
\colhead{(\%)}&
\colhead{(\%)}
}
\startdata
    1.2   &       1.1       &        1.6 \nl
    2.2   &       0.3       &        1.2 \nl
    3.5   &       0.8       &        0.2 \nl
    4.9   &       1.1       &        0.4 \nl
   12     &       0.6       &        1.1 \nl
   25     &       0.4       &        0.9 \nl
   60     &       0.1       &        1.0 \nl
  100     &      -0.3       &        1.6 \nl
  140     &      -0.7       &        1.9 \nl
  240     &      -5.4       &        1.4 \nl
\enddata
\end{deluxetable}

\end{document}


\pagestyle{empty}

\begin{deluxetable}{lllll}
\tablecolumns{5}
\tablewidth{0pc}
\tablenum{1}
\tablecaption{Density and Geometrical Parameters of the IPD Model}
\label{tbl:tabparam} 
\tablehead{
\colhead{cloud component} & 
\colhead{parameter} & 
\colhead{description} & 
\colhead{final value} &
\colhead{68\% joint. conf. unc.} }
\startdata
Smooth Cloud & $n_0$ & density at 1 AU & $1.13\times 10^{-7}$ {AU}$^{-1}$ & 
$6.4\times 10^{-10}$ \nl
(Widened     & $\alpha$ & radial power law exponent  & $1.34$  & 0.022 \nl
Modified Fan)& $\beta$  & vertical shape parameter   & $4.14$  & 0.067 \nl
             & $\gamma$ & vertical power law exponent& $0.942$ & 0.025 \nl
             & $\mu$    & widening parameter         & $0.189$ & 0.014 \nl
             & $i$      & inclination         & $2.03^\circ$   & 0.017 \nl
             & $\Omega$ & asc. node   & $77.7^\circ$           & 0.6   \nl
             & $X_0$    & x offset from Sun   & $.0119$ AU     & 0.0011\nl
             & $Y_0$    & y offset from Sun   & $.00548$ AU    & 0.00077\nl
             & $Z_0$    & z offset from Sun   & $-.00215$ AU   & 0.00043\nl
\hline
Dust Band 1 & $n_{B1}$ & density at 3 AU  & $5.59\times 10^{-10}$ {AU}$^{-1}$ &
$7.20\times 10^{-11}$\nl
            & $\delta_{\zeta_{B1}}$ & shape parameter & 8.78$^\circ$ & fixed \nl
            & $v_{B1}$              & shape parameter & .10          & fixed \nl
            & $p_{B1}$              & shape parameter & 4            & fixed \nl
            & $i_{B1}$              & inclination    & 0.56$^\circ$  & fixed \nl
            & $\Omega_{B1}$         & asc. node      & 80$^\circ$    & fixed \nl
            & $\delta_{{R}_{B1}}$  & inner radial cutoff & 1.5 {AU} & fixed \nl
\hline
Dust Band 2 & $n_{B2}$ & density at 3 AU  & $1.99\times 10^{-9}$ {AU}$^{-1}$ &
$1.28\times 10^{-10}$\nl
            & $\delta_{\zeta_{B2}}$ & shape parameter & 1.99$^\circ$ & fixed \nl
            & $v_{B2}$              & shape parameter & .90          & fixed \nl
            & $p_{B2}$              & shape parameter & 4            & fixed \nl
            & $i_{B2}$              & inclination    & 1.2$^\circ$   & fixed \nl
            & $\Omega_{B2}$         & asc. node      & 30.3$^\circ$  & fixed\nl
            & $\delta_{{R}_{B2}}$   & inner radial cutoff & .94 {AU} & 0.025\nl
\hline
Dust Band 3 & $n_{B3}$ & density at 3 AU  & $1.44\times 10^{-10}$ {AU}$^{-1}$ &
$2.34\times 10^{-11}$\nl
            & $\delta_{\zeta_{B3}}$ & shape parameter & 15.0$^\circ$ & fixed \nl
            & $v_{B3}$              & shape parameter & .05          & fixed \nl
            & $p_{B3}$              & shape parameter & 4            & fixed \nl
            & $i_{B3}$              & inclination    & 0.8$^\circ$   & fixed \nl
            & $\Omega_{B3}$         & asc. node      & 80.0$^\circ$  & fixed \nl
            & $\delta_{{R}_{B3}}$  & inner radial cutoff & 1.5 {AU} & fixed \nl
\hline
Solar Ring  & $n_{SR}$        & density at 1 AU & $1.83\times 10^{-8}$ AU$^{-1}$
& $1.27\times 10^{-9}$\nl
            & $R_{SR}$        & radius of peak density  & $1.03$ AU & 0.00016\nl
            & $\sigma_{rSR}$  & radial dispersion      & $0.025$ AU & fixed \nl
            & $\sigma_{zSR}$  & vertical dispersion    & $0.054$ AU & 0.0066\nl
            & $i_{RB}$        & inclination & $0.49^\circ$  & 0.063 \nl
            & $\Omega_{RB}$   & asc. node   & $22.3^\circ$  & 0.0014 \nl
\hline
Trailing Blob & $n_{TB}$      & density at 1 AU & $1.9\times 10^{-8}$ AU$^{-1}$
&$1.42\times 10^{-9}$\nl
            & $R_{TB}$        & radius of peak density  & $1.06$ AU & 0.011\nl
            & $\sigma_{rTB}$  & radial dispersion      & $0.10$ AU  & 0.0097\nl
            & $\sigma_{zTB}$  & vertical dispersion    & $0.091$ AU & 0.013 \nl
            & $\theta_{TB}$   & longitude w.r.t. earth & -10$^\circ$ & fixed \nl
            & $\sigma_{\theta TB}$ & longitude dispersion& $12.1^\circ$ & 3.4\nl
\hline
\enddata

\end{deluxetable}


\pagestyle{empty}

\begin{deluxetable}{lllll}
\tablecolumns{5}
\tablewidth{0pc}
\tablenum{2}
\tablecaption{Source Function Parameters of the IPD Model}
\label{tbl:tabparam} 
\tablehead{
\colhead{cloud component} & 
\colhead{parameter} & 
\colhead{description} & 
\colhead{final value} &
\colhead{68\% joint. conf. unc.} }
\startdata
All &$T_0$  & Temperature at 1 AU  & 286 K & fixed \nl
   &$\delta$ & Temperature power law exponent & 0.467 & 0.0041\nl
   &$C_{0,1}$ & $1.25 \mu$m phase func. parameter & -0.942 sr$^{-1}$ & fixed\nl
   &$C_{1,1}$ & $1.25 \mu$m phase func. parameter & 0.121 rad$^{-1}$ sr$^{-1}$ & fixed\nl
   &$C_{2,1}$ & $1.25 \mu$m phase func. parameter & -0.165 rad$^{-1}$ &fixed\nl
   &$C_{0,2}$ & $2.2 \mu$m phase func. parameter & -0.527 sr$^{-1}$& fixed\nl
   &$C_{1,2}$ & $2.2 \mu$m phase func. parameter &  0.187 rad$^{-1}$ sr$^{-1}$& fixed\nl
   &$C_{2,2}$ & $2.2 \mu$m phase func. parameter & -0.598 rad$^{-1}$& fixed\nl
   &$C_{0,3}$ & $3.5 \mu$m phase func. parameter & -0.431 sr$^{-1}$& fixed\nl
   &$C_{1,3}$ & $3.5 \mu$m phase func. parameter &  0.172 rad$^{-1}$ sr$^{-1}$& fixed\nl
    &$C_{2,3}$ & $3.5 \mu$m phase func. parameter & -0.633 rad$^{-1}$& fixed\nl
\hline
Smooth Cloud & $A_1$ & Albedo, $1.25 \mu$m & $.204$ & 0.0013\nl
             & $A_2$ & Albedo, $2.2 \mu$m  & $.255$ & 0.0017\nl
             & $A_3$ & Albedo, $3.5 \mu$m  & $.210$ & 0.019\nl
             & $E_3$ & Emissivity mod. factor, $3.5 \mu$m & $1.66$ & 0.088 \nl
             & $E_4$ & Emissivity mod. factor, $4.9 \mu$m & $0.997$ & 0.0036\nl
             & $E_5$ & Emissivity mod. factor, $12 \mu$m & $0.958$ & 0.0026\nl
             & $E_6$ & Emissivity mod. factor, $25 \mu$m & $1.00$ & fixed \nl
             & $E_7$ & Emissivity mod. factor, $60 \mu$m & $0.733$ & 0.0055\nl
             & $E_8$ & Emissivity mod. factor, $100 \mu$m & $0.647$ & 0.012 \nl
             & $E_9$ & Emissivity mod. factor, $140 \mu$m & $0.677$ & -\nl
             & $E_{10}$ & Emissivity mod. factor, $240 \mu$m & $0.519$ & - \nl
\hline
Dust Bands   & $A_1$ & Albedo, $1.25 \mu$m & $.204$ & fixed to smooth cloud \nl
             & $A_2$ & Albedo, $2.2 \mu$m  & $.255$ &fixed to smooth cloud \nl
             & $A_3$ & Albedo, $3.5 \mu$m  & $.210$ & fixed to smooth cloud \nl
             & $E_3$ & Emissivity mod. factor, $3.5 \mu$m & $1.66$ & fixed to smooth cloud \nl
             & $E_4$ & Emissivity mod. factor, $4.9 \mu$m & $0.359$& 0.054 \nl
             & $E_5$ & Emissivity mod. factor, $12 \mu$m & $1.01$ & 0.15\nl
             & $E_6$ & Emissivity mod. factor, $25 \mu$m & $1.00$& fixed \nl
             & $E_7$ & Emissivity mod. factor, $60 \mu$m & $1.25$& 0.30 \nl
             & $E_8$ & Emissivity mod. factor, $100 \mu$m & $1.52$& 0.65 \nl
             & $E_9$ & Emissivity mod. factor, $140 \mu$m & $1.13$& - \nl
             & $E_{10}$ & Emissivity mod. factor, $240 \mu$m & $1.40$& - \nl
\hline
Ring + Blob  & $A_1$ & Albedo, $1.25 \mu$m & $.204$& fixed to smooth cloud \nl
             & $A_2$ & Albedo, $2.2 \mu$m  & $.255$& fixed to smooth cloud \nl
             & $A_3$ & Albedo, $3.5 \mu$m  & $.210$& fixed to smooth cloud \nl
             & $E_3$ & Emissivity mod. factor, $3.5 \mu$m & $1.66$& fixed to smooth cloud \nl
             & $E_4$ & Emissivity mod. factor, $4.9 \mu$m & $1.06$& 0.0089\nl
             & $E_5$ & Emissivity mod. factor, $12 \mu$m & $1.06$& 0.00078\nl
             & $E_6$ & Emissivity mod. factor, $25 \mu$m & $1.00$& fixed \nl
             & $E_7$ & Emissivity mod. factor, $60 \mu$m & $0.873$& 0.0042\nl
             & $E_8$ & Emissivity mod. factor, $100 \mu$m & $1.10$& $7.5\times
10^{-6}$\nl
             & $E_9$ & Emissivity mod. factor, $140 \mu$m & $1.15$& -\nl
             & $E_{10}$ & Emissivity mod. factor, $240 \mu$m & $0.858$& - \nl
\enddata

\end{deluxetable}